\documentstyle[twocolumn,floats,aps,amsmath,latexsym,amssymb,epsfig]{revtex}

\begin{document}

\newcommand{\beeq}{\begin{equation}}
\newcommand{\ene}{\end{equation}}
\newcommand{\bea}{\begin{eqnarray}}
\newcommand{\ena}{\end{eqnarray}}
\newcommand{\no}{\noindent}
\newcommand{\nb}{\nonumber}
\newcommand{\Ra}{\Rightarrow}
\newcommand{\ra}{\rightarrow}
\newcommand{\Bigbreak}{\par \ifdim\lastskip < \bigskipamount \removelastskip
\fi \penalty-300 \vskip 10mm plus 5mm minus 2mm}

\newcommand{\1}{{\bf 1}}
\newcommand{\e}{{\rm e}}
\newcommand{\Hi}{{\cal H}}
\newcommand{\D}{{\cal D}}
\newcommand{\Li}{{\cal L}}
\newcommand{\U}{{\cal U}}
\newcommand{\E}{{\cal E}}
\newcommand{\F}{{\cal F}}
\newcommand{\G}{{\cal G}}
\newcommand{\Pa}{{\cal P}}
\newcommand{\V}{{\cal V}}
\newcommand{\W}{{\cal W}}
\newcommand{\Si}{{\cal S}}
\newcommand{\M}{{\cal M}}
\newcommand{\A}{{\cal A}}
\newcommand{\B}{{\cal B}}
\newcommand{\Oa}{{\cal O}}
\newcommand{\Q}{{\cal Q}}
\newcommand{\T}{{\cal T}}
\newcommand{\cZ}{{\cal Z}}

\newcommand{\R}{\Bbb{R}}

\newcommand{\der}{\partial }
\newcommand{\mis}{\frac{dk}{ 2 \pi }}

\newcommand{\sci}{\varphi }
\newcommand{\dsc}{\widetilde \varphi}
\newcommand{\scr}{\varphi_{{}_R}}
\newcommand{\scl}{\varphi_{{}_L}}
\newcommand{\scz}{\varphi_{{}_Z}}
\newcommand{\sczu}{\varphi_{{}_{Z_1}}}
\newcommand{\sczd}{\varphi_{{}_{Z_2}}}
\newcommand{\dlt}{\delta_{{}_{Z_1}{}_{Z_2}}}
\newcommand{\pr}{\varphi^+_{{}_R}}
\newcommand{\mr}{\varphi^-_{{}_R}}
\newcommand{\ml}{\varphi^-_{{}_L}}
\newcommand{\pz}{\varphi^+_{{}_Z}}
\newcommand{\mz}{\varphi^-_{{}_Z}}
\newcommand{\fz}{f_{{}_Z}}
\newcommand{\jz}{j_{{}_Z}}
\newcommand{\az}{\A_{{}_Z}}
\newcommand{\Qr}{Q_{{}_R}}
\newcommand{\Ql}{Q_{{}_L}}
\newcommand{\Qz}{Q_{{}_Z}}
\newcommand{\qr}{q_{{}_R}}
\newcommand{\ql}{q_{{}_L}}
\newcommand{\qz}{q_{{}_Z}}
\newcommand{\cqr}{{\Q}_{{}_R}(f_{{}_R})}
\newcommand{\cql}{{\Q}_{{}_L}(f_{{}_L})}
\newcommand{\cqz}{{\Q}_{{}_Z}(f_{{}_Z})}
\newcommand{\cqzu}{{\Q}_{{}_{Z_1}}(f_{{}_{Z_1}})}
\newcommand{\cqzd}{{\Q}_{{}_{Z_2}}(f_{{}_{Z_2}})}
\newcommand{\spr}{\psi_{{}_R}}
\newcommand{\spl}{\psi_{{}_L}}
\newcommand{\ep}{{\e^{i|k|x^0-ikx^1}}}
\newcommand{\emi}{{\e^{-i|k|x^0+ikx^1}}}
\newcommand{\muz}{\mu_{{}_Z}}

\newcommand{\alg}{\cal A}
\newcommand{\palg}{{\cal A}_\lambda }
\newcommand{\form}{\langle \, \cdot \, , \, \cdot \, \rangle }
\newcommand{\sform}{(\, \cdot \, ,\, \cdot \, )}
\newcommand{\Hl}{\overline \R_+}
\newcommand{\cotanh}{{\rm cotanh}}
\newcommand{\supp}{{\rm supp}}

\tightenlines

\draft

\twocolumn[\hsize\textwidth\columnwidth\hsize\csname
@twocolumnfalse\endcsname

\title{Anyon Condensation and Persistent Currents}
\author{Antonio Liguori${}^1$, Mihail Mintchev${}^{1, 2}$ 
and Luigi Pilo${}^{2, 3}$}
\address{${}^1$Dipartimento di Fisica dell'Universit\`a di Pisa, 
Via Buonarotti 2, 56127 Pisa, Italy \\ 
${}^2$Istituto Nazionale di Fisica Nucleare, Sezione di Pisa, Italy \\ 
${}^3$Scuola Normale Superiore, Piazza dei Cavalieri 7, 56126 Pisa, Italy}

\maketitle
\begin{abstract}
Condensation and persistent currents in 1+1 dimensional anyon systems are 
discovered. 
\end{abstract} 
\vskip 0.5 truecm
]

The recent interest in quantum many body systems in one space dimension
is essentially related to some applications in modern
condensed matter physics. Among others, they include
fractional quantum Hall effect \cite{L}, high temperature
super-conductivity \cite{A}, quasi-one-dimensional organic metals,
quantum dots, quantum wires, etc. Some of the just mentioned systems
exhibit properties, which require new physical concepts. 
Signatures of generalized, anyonic statistics (see e.g. \cite{W} and 
references therein) have been observed for instance in
experiments with quantum Hall bars and in one-dimensional organic metals.
This letter represents a short account of our recent results \cite{LMP}
concerning the behaviour of anyon fields at finite temperature and density.
The novel features we report on, are the existence of anyon condensation and
the presence of persistent currents. Our main tool is an operator formalism
for bosonization at finite temperature and density. We avoid the delicate problems 
related with the thermodynamic limit in presence of anyons, working directly in 
infinite volume. 

Let us summarize first the basic points of our approach. 
We start with a free massless scalar field $\sci $ and its dual 
$\dsc $ in 1+1 dimensions, which satisfy the equations of 
motion
\beeq
g^{\mu \nu}\der_\mu \der_\nu \, \sci (x) =
g^{\mu \nu}\der_\mu \der_\nu \, \dsc (x) = 0 \, ,
\label{eqm}
\end{equation}
($g = {\rm diag}\, \, (1,\, -1)$), the duality constraint
\beeq
\der_\mu \dsc (x) = \varepsilon_{\mu \nu }\, \der^\nu \sci (x) \, ,
\label{dual}
\end{equation}
($\varepsilon_{01} = 1$) and the standard equal-time canonical commutation 
relations. As it is well known, all these requirements are fulfilled by 
\bea 
&& \sci (x) = \frac{1}{2}[\scr (x^-) + \scl (x^+)] , \\
&& \dsc (x)= \frac{1}{2}[\scr (x^-) - \scl (x^+)] \, , 
\ena
where $x^{\pm} = x^0 \pm x^1$ and  
\bea
&&\scr (\zeta) \equiv
{\sqrt 2} \int_0^\infty \mis \left [a^\ast (k) \, \e^{ik\zeta } +
a(k) \, \e^{-ik \zeta } \right ] \, , \label{rf} \\
&& \scl (\zeta) \equiv
{\sqrt 2} \int^0_{-\infty }\mis \left [a^\ast (k) \, \e^{-ik\zeta } +
a(k) \, \e^{ik\zeta }\right ] \, , 
\label{lf}
\ena 
with
\bea
&&[a(k)\, ,\, a(p)] = [a^\ast (k)\, ,\, a^\ast (p)] = 0 \quad , \label{aa} \\
&& [a(k)\, ,\, a^\ast (p)] = |k|^{-1}_\lambda \, 2\pi \delta (k-p)\, 
\quad . \label{aas}
\ena
The distribution $|k|^{-1}_\lambda $, whose explicit form is not essential for 
our discussion, satisfies $|k| |k|^{-1}_\lambda  = 1 $.
The parameter $\lambda > 0$ has a well understood infrared origin \cite {Wi}. From 
Eqs. (\ref{rf}-\ref{aas}) one gets 
\beeq
[\sczu (\zeta_1 )\, ,\, \sczd (\zeta_2 )] =
-i\, \varepsilon (\zeta_{12})\, \dlt \, ,
\label{cz}
\end{equation}
where $Z = L, R$, $\; \zeta_{12} \equiv \zeta_1 - \zeta_2 $ 
and
\beeq
\dlt = \left\{ \begin{array}{cc} 1\, & \mbox{if } Z_1=Z_2 \, , \\
0\, & \mbox{if } Z_1 \neq Z_2 \, . \nb 
\end{array} \right. 
\end{equation}
We denote by $\az$ the algebra generated by 
$\scz $ and observe that 
\beeq
\alpha_{\mu_Z}\, :\, \scz (\zeta ) \longmapsto
\scz (\zeta ) - \frac{1}{\sqrt{\pi}} \, \mu_{{}_Z}\zeta \, ,
\quad \mu_{{}_Z} \in \R \, , \label{saut}
\end{equation}
is an automorphism of $\az$. The parameters $\mu_{{}_Z}$ will play the role of 
chemical potentials associated with the chiral charges 
\beeq 
q_{{}_Z} = \frac{1}{2}\int_{-\infty }^\infty d\zeta\, j_{{}_Z} (\zeta ) \, ,
\quad    \jz (\zeta ) = \frac{1}{ \sqrt \pi}\, \der \scz (\zeta )  \, .\label{chd} 
\end{equation} 
Eq. (\ref{cz}) implies  
\beeq
[q_{{{}_Z}_1}  \, , \, \sci_{{{}_Z}_2} (\zeta )] = \frac{1}{i\sqrt{\pi}}  \, \dlt \, ,
\quad  [q_{{{}_Z}_1}  \, , \, q_{{{}_Z}_2}] = 0 \, . 
\label{char1} 
\end{equation}

We now turn to generalized statistics, which can be introduced  
at purely algebraic level. In fact, let us consider the set of
fields parametrized by $\xi= (\sigma,\, \tau ) \in {\R}^2$ and defined by
\bea
&&A(x;\xi) \equiv
z(\lambda ;\xi)\, \exp \left [\frac{i\pi} {2} \left (\tau \qr - \sigma \ql
\right )\right ] \times \nb \\
&&:\exp \left \{i\sqrt \pi \left [\sigma \scr (x^-) +
\tau \scl (x^+) \right ]\right \}: \, ,
\label{adef}
\ena
where $z(\lambda ;\xi)$ is a suitable normalization factor and 
the normal ordering $\, : \quad  :\, $ is taken with respect to
$\{a^*(k), a(k) \}$. 
Using Eqs. (\ref{cz},\ref{char1}), one finds for space-like separated 
points ($x_{12}^2 < 0$) 
\bea
&&A(x_1;\xi) A(x_2;\xi) = \nb \\ 
&&\exp [-i\pi (\sigma^2 - \tau^2) \varepsilon (x^1_1-x^1_2)]
A(x_2;\xi) A(x_1;\xi) \, .
\label{exc}
\ena
Therefore, $A(x;\xi)$ is an anyon field whose statistics parameter \cite{W} is 
\beeq
\vartheta (\xi ) = \sigma^2 - \tau^2 \, . 
\end{equation} 
Bose or Fermi statistics are recovered when $\vartheta $ is an even or odd
integer respectively. The remaining values of $\vartheta $ lead to
Abelian braid statistics. What is crucial in implementing of 
generalized statistics is the relative non-locality of $\sci $ and $\dsc $. 

Let us concentrate on the family $\{A(x;\xi )\, :\, \xi \in \Xi \}$ with 
$\Xi = \{\xi_1, -\xi_1,...,\xi_p, -\xi_p \}$. In \cite{LMP} we have constructed 
a representation $\T(\Xi; \mu_{{}_L}, \mu_{{}_R})$ 
describing these fields in thermal equilibrium with inverse temperature $\beta$ and 
chemical potentials $\mu_{{}_R}, \mu_{{}_L}$. The construction is in two 
steps. Using the conventional thermal representation \cite{BR} of the 
commutation relations (\ref{aa},\ref{aas}) and the automorphism $\alpha_{\mu_{{}_Z}}$, 
we first derive a thermal representation $\T_{{}_Z}(\mu_{{}_Z})$ of $\az$.
The representation  $\T_{{}_Z}(\mu_{{}_Z})$ has indefinite metric and depends 
on the parameter $\lambda$ and a bosonic
chemical potential $\mu_{{}_B}$, introduced for technical reasons. 
The second step consists in determining 
$\T(\Xi; \mu_{{}_L}, \mu_{{}_R}) \subset \T_{{}_L}(\mu_{{}_L}) \otimes \T_{{}_R}(\mu_{{}_R})$ 
by appropriate selection rules. Referring for all details to \cite{LMP}, 
we stress that $\T(\Xi; \mu_{{}_L}, \mu_{{}_R})$ is both $\lambda$- and 
$\mu_{{}_B}$-independent and has positive metric. The basic correlation functions 
characterizing this physical representation are  
\bea
&&\langle A(x_1;\xi_1)\cdots A(x_n;\xi_n) \rangle_{\mu_L,\, \mu_R}^\beta =
\left (\frac{1}{ 2 \pi} \right )^{\frac{1}{2}\sum_{i=1}^n (\sigma_i^2 + 
\tau_i^2)} \times \nb \\
&& \exp \left [\frac{i \pi}{2} \sum_{\stackrel{i,j=1}{i<j}}^n (\tau_i \sigma_j
 - \tau_j \sigma_i )
-i\mu_{{}_R}\sum_{i=1}^n \sigma_i x_i^- - i\mu_{{}_L} 
\sum_{i=1}^n \tau_i x_i^+ \right ] 
\nb \\
&&
\prod_{\stackrel{i,j=1}{ i<j}}^n
\left [ \frac{i \beta}{ \pi } \sinh
\left (\frac{\pi}{ \beta} x_{ij}^- -i\epsilon \right )\right ]^{\sigma_i
\sigma_j} \negthickspace  \negthickspace
\left [ \frac{i \beta}{  \pi } \sinh
\left (\frac{\pi }{ \beta}x_{ij}^+ -i\epsilon \right )\right ]^{\tau_i \tau_j}
\negthickspace  \negthickspace \negthickspace  \negthickspace
\negthickspace  \negthickspace \negthickspace  \negthickspace
\label{ac}
\ena
where $\xi_i \in \Xi$ and 
\beeq 
\sum_{i=1}^n \sigma_i = \sum_{i=1}^n \tau_i = 0 \, . 
\label{sc}
\end{equation}
All correlators violating (\ref{sc}) vanish, whereas the 
insertion of $m$ current operators $ j_{{}_{Z_j}}$ in (\ref{ac})
is obtained by iteration, using
\bea
&&\langle j_{{}_{Z_1}}(\zeta_1)j_{{}_{Z_2}}(\zeta_2)
\cdots j_{{}_{Z_m}}(\zeta_m)A(x_1;\xi_1)\cdots  A(x_n;\xi_n )
\rangle_{\mu_{{}_L},\, \mu_{{}_R}}^\beta \nb \\
&&= -\left \{ \frac{\mu_{{}_{Z_1}} }{ \pi} + \frac{i}{ \beta} \sum_{j=1}^n
\xi_j^{Z_j}
\coth \left [ \frac{\pi}{ \beta}(\zeta_1 - x_j^{Z_1}) - i\epsilon \right ]
\right \} \nb \\
&&
\langle j_{{}_{Z_2}}(\zeta_2)\cdots
j_{{}_{Z_m}}(\zeta_m)A(x_1;\xi_1)\cdots  A(x_n;\xi_n )
\rangle_{\mu_{{}_L},\, \mu_{{}_R}}^\beta \nb \\
&&
- \frac{1}{ \beta^2}\sum_{j=2}^m \delta_{{}_{Z_1}{}_{Z_j}} \sinh^{-2}\left 
(\frac{\pi}{\beta }
\zeta_{1j} - i\epsilon \right )  
\langle j_{{}_{Z_2}}(\zeta_2)\cdots \widehat {j_{{}_{Z_j}}}(\zeta_j)
\nb \\
&&
\cdots j_{{}_{Z_m}}(\zeta_m)A(x_1;\xi_1)\cdots  A(x_n;\xi_n )
\rangle_{\mu_{{}_L},\, \mu_{{}_R}}^\beta  \, ,
\label{jac}
\ena 
where
\beeq
\xi^Z = \left\{ \begin{array}{cc} \sigma  & \mbox{if } Z=R\, , \\
\tau  & \mbox{if } Z=L\, , \end{array} \right. \qquad
x^Z =  \left\{ \begin{array}{cc} x^- & \mbox{if } Z=R\, ,  \\
x^+  & \mbox{if } Z=L\, , \end{array} \right. \nb 
\end{equation}
and the hat in the right hand side of (\ref{jac}) indicates that the
corresponding current must be omitted. Eq. (\ref{jac}) implies
\beeq
\langle j_{{}_Z} (\zeta) \rangle_{\mu_{{}_L},\, \mu_{{}_R}}^\beta = 
- \frac{\mu_{{}_Z}}{\pi} \, , 
\end{equation} 
which is the origin of non-vanishing charge density and persistent current. 

The explicit form of the correlators is quite
remarkable and deserves a comment. Without current insertions, 
the equal-time $n$-point function (\ref{ac}) is
a finite temperature and density generalization of the Jastrow-Laughlin
wave function \cite{L}. We recall that the latter  describes the 
$n$-particle ground state in several
one-dimensional models, showing a Tomonaga-Luttinger liquid structure. In that 
context, the current insertions in (\ref{ac}) are associated with the charged
excitations of the liquid. The expectation values (\ref{ac},\ref{jac}) are invariant 
under the transformations:
\bea
&&A(x ;\xi ) \rightarrow A(x^0+t ,x^1 ;\xi )\, , \, \; \quad
 j_{{}_Z}(\zeta ) \rightarrow j_{{}_Z}(\zeta + t)\, , \nb \\
&&A(x;\xi ) \rightarrow \e^{-is_{{}_R}\sigma }\, A(x;\xi )
\, , \quad \quad j_{{}_Z}(\zeta ) \rightarrow j_{{}_Z}(\zeta )\, ,  \nb \\
&&A(x;\xi ) \rightarrow \e^{-is_{{}_L}\tau }\, A(x;\xi ) \, ,
 \quad \quad j_{{}_Z}(\zeta ) \rightarrow j_{{}_Z}(\zeta )\, , \nb
\ena
with $t, \, s_{{}_Z} \in \R$. If one 
denotes the corresponding conserved charges by $H$ and $\Qz$ and 
considers the automorphism $\alpha_s$, generated by
\beeq
K \equiv H -\mu_{{}_L}\Ql - \mu_{{}_R}\Qr \, ,
\end{equation}
one can prove \cite{LMP} that 
the correlation functions (\ref{ac},\ref{jac}) obey the
Kubo-Martin-Schwinger condition relative to $\alpha_s$ with (inverse) 
temperature $\beta $. Clearly, this result is fundamental for the 
physical interpretation of the representation $\T(\Xi; \mu_{{}_L}, \mu_{{}_R})$. 

The equations of motion, induced by the correlation functions (\ref{jac}), are 
\beeq
i \pi \xi^Z\, \vdots \, j_{{}_Z}(x^Z)A(x;\xi )\, \vdots =
\frac{\der}{ \der x^Z } A(x;\xi )\, . 
\label{ja}  
\end{equation}
where the normal product $\vdots \, \cdots \, \vdots $ is defined by 
\bea
&&\vdots \, j_{{}_Z}(x^Z)A(x;\xi )\, \vdots \equiv 
\lim_{y\to x} \bigl\{ j_{{}_Z}(y^Z)A(x;\xi ) \nb  \\
&& \left. +  \frac{i\xi^Z}{ \beta} \coth \left [\frac{\pi}{  \beta }(y-x)^Z - i
\epsilon \right ] A(x;\xi ) \right \} \, . \nb  
\ena 
The chiral components of the energy-momentum tensor are
\beeq
\Theta_{{}_Z} (\zeta) =
\frac{1}{ 4}\lim_{\zeta'\to\zeta}\left[\pi \jz (\zeta') \jz(\zeta)
+ \frac{1}{ \pi (\zeta'-\zeta-i\epsilon)^2}\right] \, . \nb 
\end{equation}
Expressed in terms of $\Theta_{{}_Z}$, the Hamiltonian $H$ reads
\beeq
H = \int_{-\infty }^\infty  d\zeta \, \left[\Theta_{{}_R}(\zeta) +  
\Theta_{{}_L}(\zeta) \right] \, . \nb 
\end{equation}

In order to get a deeper insight into the physical properties of the field
$A(x;\xi )$, it is instructive to derive the relative momentum 
distribution. For this purpose we consider the Fourier transform 
\beeq
{\widehat W}^\beta (\omega , k;\xi ) = \int_{-\infty }^\infty 
\negthickspace  \negthickspace \negthickspace  \negthickspace d^2x  
\e^{i\omega x^0 - ikx^1}  \langle A^*(x_1;\xi)A(x_2;\xi) 
\rangle_{\mu_L,\, \mu_R}^\beta . \nb   
\end{equation}
Inserting the explicit expression (see Eq. (\ref{ac})) of the 
two-point function, one gets 
\bea
&&{\widehat W}^\beta (\omega , k;\xi ) = \frac {1}{2} \times \nb \\
&&\varrho \left (\frac{\omega }{2} + \frac{k}{2} + 
\mu_{{}_R} \sigma ;\, \sigma^2 , \beta \right ) 
\varrho \left (\frac{\omega }{2} - \frac{k}{2} + 
\mu_{{}_L} \tau ;\, \tau^2 , \beta \right ) \, , 
\label{roro} 
\ena
with 
\bea 
&&\varrho (k ;\alpha ,\, \beta ) =  \negthickspace  \negthickspace 
\negthickspace
\negthickspace \negthickspace  \negthickspace \negthickspace  \negthickspace
\label{dis} \\
&&\left \{ \begin{array}{cc}(\beta )^{1-\alpha}
\, \e^{\frac{1}{ 2}\beta k }\, \Big| \Gamma \left (\frac{1}{  2}\alpha +
\frac{i}{ 2\pi }\beta k \right ) \Big|^2 [2\pi \Gamma (\alpha )]^{-1}& 
\mbox{for } \alpha > 0  \\ 
2 \pi \, \delta (k) & \mbox{for } \alpha = 0 .  
\end{array}  \right.  
\nb
\ena
Observing that the contributions of the left- and right-moving modes factorize, 
we consider first the cases 
\beeq 
{\widehat W}^\beta (\omega , k;(\sigma ,0)) = 
2\pi \delta (\omega - k) 
\varrho \left (\mu_{{}_R}\sigma + k;\, \sigma^2 , \beta \right ) \, , 
\label{rro} 
\end{equation}
and 
\beeq 
{\widehat W}^\beta (\omega , k;(0,\tau )) = 
2\pi \delta (\omega + k) 
\varrho \left (\mu_{{}_L}\tau - k;\, \tau^2 , \beta \right ) \, . 
\label{lro} 
\end{equation}
Eqs. (\ref{rro},\ref{lro}) have a simple physical interpretation: 
the $\delta $-factors fix the dispersion 
relations, whereas the $\varrho $-factors give the momentum distributions. From 
the exchange relation (\ref{exc}) we already know  
that the fields $A(x;(\pm 1,0))$ and $A(x;(0,\pm 1))$ have Fermi statistics. 
In agreement with this fact, one gets from (\ref{rro},\ref{lro}) 
\beeq 
{\widehat W}^\beta (\omega , k;(\pm 1,0)) = 
2\pi \delta (\omega - k) \, \frac{1}{ 1+\e^{-\beta (k\, \pm \, \mu_{{}_R})}}\, , 
\end{equation}
\label{rdd}
\beeq
{\widehat W}^\beta (\omega , k;(0,\pm 1)) = 
2\pi \delta (\omega + k)\, \frac{1}{ 1+\e^{\beta (k\, \mp \,\mu_{{}_L})}} \, .
\label{ldd}
\end{equation}
We thus recover the familiar Fermi distribution (with Fermi momenta 
$k_{{}_F} = \mp \, \mu_{{}_R}$ and $k_{{}_F} = \pm \, \mu_{{}_L}$), which 
represents an useful check. 
\begin{figure}[h]
\begin{picture}(13,13)
\put(220,-125){$ k$}
\put(120, 0){$\varrho$}
\end{picture}
\vskip -1.  truecm 
\centerline{\hskip 1truecm \epsfig{file=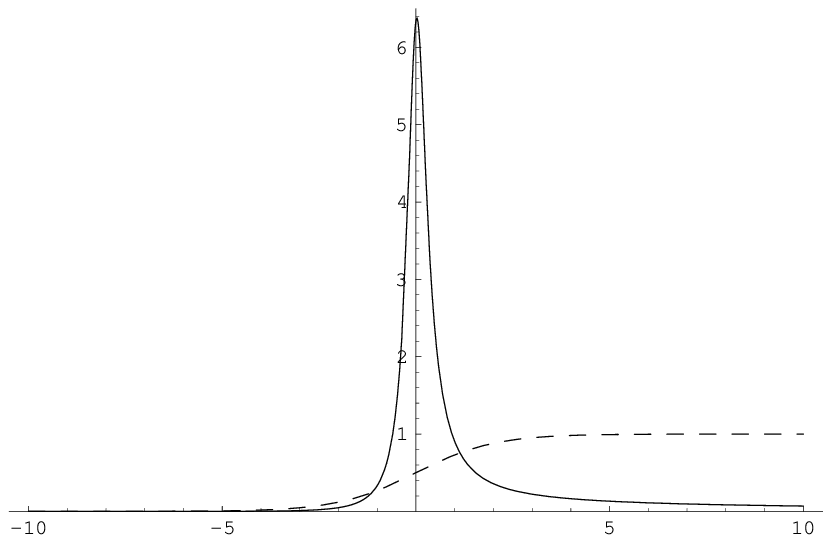,height=13cm,
width=13cm}}
\vskip -7.5 truecm 
\noindent FIG. 1. The distribution $\varrho (k; \alpha , \beta=1)$ for 
$\alpha = 1$ (dashed line) and $\alpha = 0.1$ (continuous line). 
\vskip -0.5 truecm 
\end{figure} 
Turning back to the general expression (\ref{dis}), we see that for 
$\alpha >0$ 
the distribution $\varrho $ is a smooth positive function, whose asymptotic 
behaviour is encoded in
\bea
&& \varrho (k;\alpha ,\beta ) \sim \frac{1}{  \Gamma (\alpha )} 
\left (\frac{k}{  2\pi}\right )^{\alpha -1} \, , \, \, \, \, 
\qquad k\to \infty \, , 
\label{kp}\\
&&\varrho (k;\alpha ,\beta ) \sim \frac{e^{\beta k}}
{ \Gamma (\alpha )}\left (-\frac{k}{ 2\pi}\right )^{\alpha -1} , 
\qquad k\to -\infty \, . 
\label{km}
\ena
In the range $\alpha \geq 1$, $\varrho $ is monotonically increasing on the 
whole line $k \in \R$. When $0 < \alpha < 1$, $\varrho $ increases monotonically 
for $k\leq 0$ and, according to Eqs. (\ref{kp},\ref{km}), admits at least one 
local maximum for $k > 0$. Let us denote the position of the first one (when $k$ 
moves from $0$ to $\infty $) by $k_{{}_C}(\alpha, \beta )$. It is easily seen that 
$k_{{}_C}(\alpha, \beta ) = \beta^{-1} s_{{}_C}(\alpha )$, where 
$s_{{}_C}(\alpha )$ is a solution of certain functional equation, 
which is not displayed for the sake of conciseness. We have some numerical 
evidence that the maximum $k_{{}_C}(\alpha, \beta )$ is unique.
The plots of $\varrho $ in Fig. 1 indicate 
an interesting condensation-like behaviour around $k_{{}_C}$. For comparison we 
have plotted there for $\beta =1$ the cases $\alpha = 1$ (Fermi distribution) and 
$\alpha = 0.1$. The phenomenon is clearly 
marked for small values of the temperature and/or of the parameter 
$\alpha $ in the domain $0 < \alpha < 1$. We find quite 
remarkable that for any fixed temperature, one can achieve an arbitrary sharp 
anyon condensation, taking a sufficiently small value of $\alpha $.
Concerning the behaviour of ${\widehat W}^\beta (\omega , k;\xi )$ 
when both $\sigma \not=0$ and $\tau \not= 0$, 
the above analysis implies condensation at 
\beeq 
\omega = k_{{}_C}(\sigma^2, \beta) + k_{{}_C}(\tau^2, \beta) - 
\mu_{{}_R}\sigma - \mu_{{}_L}\tau \, , 
\end{equation}
\beeq 
k = k_{{}_C}(\sigma^2, \beta) - k_{{}_C}(\tau^2, \beta) - 
\mu_{{}_R}\sigma + \mu_{{}_L}\tau \, , 
\end{equation}
provided that $0<\sigma^2 <1$ and $0<\tau^2 <1$ 
We stress that the condensation effect we discovered, does not 
contradict the Hohenberg-Mermin-Wagner (HMW) theorem on the absence of 
condensation in one space dimension. The point is that 
anyonic statistics can be equivalently described by a suitable exchange 
interaction with two- and three-body potentials, determined \cite{LiMi} 
by the corresponding exchange factor. When these potentials are confining, 
some assumptions of the HMW theorem are violated and condensation may occur 
\cite{BK}, \cite{No} even in one dimension. 
Summarising, we have shown that the right (left) moving modes of the anyon 
field $A(x;\xi )$ condensate in the range $0<\sigma^2 < 1$ ($0<\tau^2 < 1$). 

The thermal representations $\T(\Xi; \mu_{{}_L}, \mu_{{}_R})$ have universal 
character. As a concrete application, let us consider the 2-d Thirring model.
Classically, the model describes a two-component field $\Psi$ satisfying the 
equation of motion
\beeq
i\gamma^\nu \der_\nu \Psi (x) = g\pi\, [{\overline \Psi}(x)\gamma_\nu \Psi
(x)] \gamma^\nu \Psi (x) \, ,
\label{them}
\end{equation}
where $g \in \R$ is the coupling constant and
\beeq
\Psi (x)= \left( \begin{array}{c} \Psi_1(x) \\ \Psi_2(x) \end{array} \right), 
\; \gamma^0=  \left( \begin{array}{cc} 0 & 1 \\ 1 & 0 \end{array} \right),
\; \gamma^1=  \left( \begin{array}{cc} 0 & -1 \\ 1 & 0 \end{array} \right). \nb 
\end{equation}
Both the vector and the chiral current,
\beeq
J_\nu (x) = {\overline \Psi}(x)\gamma_\nu \Psi (x), \quad
J^5_\nu (x) = {\overline \Psi}(x)\gamma_\nu \gamma^5 \Psi (x),
\label{thc}
\end{equation}
($\gamma^5 = \gamma^0 \gamma^1$) are conserved and  satisfy the duality 
relation
\beeq
J^5_\nu (x) =\varepsilon_{\nu \mu }J^\mu (x) \, .
\label{thdual}
\end{equation}

The problem of quantizing this system at finite temperature and
chemical potentials $\mu $ and $\mu^5$, associated with the charges $Q$ and 
$Q^5$ and generated by the currents (\ref{thc}), has been solved in 
\cite{LMP}. We have constructed there an anyonic solution (see also 
\cite{IT}) with statistical parameter $\vartheta > -g$. The basic building 
block of our construction is the representation 
$\T(\Xi_{{}_T}; \mu_{{}_L}, \mu_{{}_R})$ where 
$\Xi_{{}_T} = \{\xi,-\xi, \xi^\prime, -\xi^\prime \}$ with 
$\xi = (\sigma ,\tau )$, $\xi^\prime = (\tau, \sigma )$ and 
\beeq
\mu_{{}_R} = \frac{1}{ \sigma + \tau }\mu + \frac{1}{ \sigma - \tau }\mu_{{}_5}
\, , \; 
\mu_{{}_L} = \frac{1}{ \sigma + \tau }\mu - \frac{1}{ \sigma - \tau }\mu_{{}_5}
\, .
\label{y}
\end{equation}
It turns out that $\Psi_1(x) = A(x;\xi )$ and $\Psi_2(x) = A(x;\xi^\prime )$, 
the quantum counterpart of Eq. (\ref{them}) being Eq. (\ref{ja}). Finally, 
in terms of $g$ and $\vartheta $ one has 
\beeq
\sigma = \pm \frac{g+2\vartheta }{2\sqrt {g+\vartheta}} , \quad 
\tau = \pm \frac{g}{2\sqrt {g+\vartheta}} \, . 
\end{equation} 
Referring again for details to \cite{LMP}, we observe that 
the correlation functions of the Thirring field $\Psi $ and of both vector and
and axial currents follow directly from 
Eqs. (\ref{ac},\ref{jac}). In particular, for the expectation value of the 
current $J_\nu (x)$ one finds
\beeq
\langle J_0(x)\rangle_{\mu,\, \mu_{{}_5}}^\beta =
\frac{\mu}{  \pi (g + \vartheta )} \, , \qquad
\langle J_1(x)\rangle_{\mu,\, \mu_{{}_5}}^\beta =
- \frac{\mu_{{}_5}}{  \pi \vartheta } \, .
\label{gibbs}
\end{equation} 
The expression for the charge density resolves some discrepancies present in 
the literature, confirming the result of \cite{sp} and extending it to the case 
$\vartheta \not= 1$. The appearance of a persistent current
$\langle J_1(x)\rangle_{\mu,\, \mu_{{}_5}}^\beta $ in the finite temperature
Thirring model represents to our knowledge a novel feature. Let us recall in 
this respect that persistent currents of quantum origin have been experimentally 
observed (\cite{LDDB}, \cite{MCB}) in mesoscopic rings placed in an external 
magnetic field. Such fields are absent in the two-dimensional world, but 
chiral symmetry, combined with duality still allow for a non-vanishing 
$\langle J_1(x)\rangle_{\mu,\, \mu_{{}_5}}^\beta $. Notice that the 
persistent current grows like $\vartheta^{-1}$ for small values of $\vartheta $. 

Concerning the energy-momentum tensor $T_{\mu \nu}(x)$ of the Thirring model,
we find that
\bea
&&  T_{00}(x)  =    T_{11}(x) \equiv \Theta_{{}_L}(x^+)+\Theta_{{}_R}(x^-)
\, , \nb \\
&&T_{01}(x) =   T_{10}(x) \equiv \Theta_{{}_L}(x^+)
-\Theta_{{}_R}(x^-)\, , \nb 
\ena
which lead to the energy and momentum densities 
\bea
&& \langle  T_{00}(x) \rangle_{\mu,\, \mu_{{}_5}}^\beta
= \frac{\pi}{6 \beta^2} + \frac{g+ \vartheta }{2\pi } 
\left [ \frac{ \mu^2 }{(g + \vartheta )^2} + 
\frac{\mu_{{}_5}^2}{\vartheta^2}  \right] \, , \nb \\
&&\langle  T_{01}(x) \rangle_{\mu,\, \mu_{{}_5}}^\beta
= -\frac{\mu \mu_{{}_5}}{\pi \vartheta } \, , \nb
\ena
and determine the equation of state \cite{LMP}. 

In conclusion, we have shown that finite temperature and density anyon systems 
in 1+1 dimensions exhibit condensation and persistent currents. These phenomena 
deserve further attention both from the conceptual and applicative point of 
view.

\end{document}